\documentclass[conference]{IEEEtran}
\usepackage{cite}
\usepackage{amsmath,amssymb,amsfonts}
\usepackage{algorithmic}
\usepackage{graphicx}
\usepackage{textcomp}
\usepackage{xcolor}

\usepackage[absolute]{textpos}
\usepackage{subcaption}

\def\BibTeX{{\rm B\kern-.05em{\sc i\kern-.025em b}\kern-.08em
    T\kern-.1667em\lower.7ex\hbox{E}\kern-.125emX}}
\begin{document}

\begin{textblock}{15}(2.5,.25)
\noindent This work has been accepted for publication at the Conference on Networked Systems (NetSys 2021).
\end{textblock}

\title{Zero Trust Service Function Chaining}

\author{\IEEEauthorblockN{Leonard Bradatsch}
\IEEEauthorblockA{ Institute of Distributed Systems\\
Ulm University\\
Ulm, Germany\\
Email: leonard.bradatsch@uni-ulm.de}
\and
\IEEEauthorblockN{Frank Kargl}
\IEEEauthorblockA{ Institute of Distributed Systems\\
Ulm University\\
Ulm, Germany\\
Email: frank.kargl@uni-ulm.de}
\and
\IEEEauthorblockN{Oleksandr Miroshkin}
\IEEEauthorblockA{ Communication and Information Center\\
Ulm University\\
Ulm, Germany\\
Email: oleksandr.miroshkin@uni-ulm.de}}

\maketitle

\begin{abstract}
In this paper, we address the inefficient handling of traditional security functions in Zero Trust (ZT) networks.
For this reason, we propose a novel network security concept that combines the ideas of ZT and Service Function Chaining (SFC). This allows us to efficiently decide which security functions to apply to which packets and when.
\end{abstract}


\section{Introduction}
ZT~\cite{b1} is one of the most discussed trends in the area of network security. The idea of ZT is motivated by current security challenges such as the trend to home work that results in new resource access patterns. ZT mainly addresses these challenges by strictly enforced trust-based authentication and authorization (Auth*) for all resource requests. The most common ZT component that enforces such Auth* decisions is the centrally positioned Policy Enforcement Point (PEP), which must be passed before resource access is granted. In current ZT architectures, traditional security functions such as Multi-factor Authentications (MFA) or Intrusion Prevention Systems (IPS) work independently from the ZT components and are applied to all packets that flow through these functions' paths. Likewise, these functions are still necessary because they provide elementary security functionalities such as DDoS protection that are not provided by the ZT compotents.
However, in modern networks with ever increasing bandwidths, the processing of all packets by such computationally expensive security functions causes a high load on the hardware that hosts the functions and may lead eventually to congestion at these points. To address this problem, we propose a dynamic orchestration of the traditional security functions in the network by the PEP. This allows the PEP to efficiently decide
when to apply which traditional security function to which packet flows. For example, an IPS can be saved if the requesting device is already sufficiently trusted after the PEP's Auth* decision to access the resource.\\
In the remainder of this paper, we introduce a novel security concept that implements the described problem solution by combining the ideas of ZT and SFC~\cite{b2}. Furthermore, we present a first prototype based on TLS and HTTP. 
As a conclusion, we give a first resume as well as an impression of our upcoming works in this area.

\section{Zero Trust Service Function Chaining}
To achieve a dynamic orchestration of traditional security functions by the ZT components and thus a more efficient security-relevant packet processing, we propose a novel network security concept called Zero Trust Service Function Chaining (ZTSFC). In this paper, we focus on an implementation of the presented ZTSFC concept in LANs such as enterprise networks. \par SFC~\cite{b2} alone describes the approach of dynamically chaining a set of available network functions (e.g., HTTP header enrichers, Congestion Mitigators or Firewalls) based on a decision made by a central classifier. Such a decision is based on certain parameters such as IP addresses that packets of a network flow have. All thus chained functions are then applied to the packets of that very flow. \par On the other hand, the most central components in a ZT architecture are the Auth* decision making Policy Decision Point (PDP) and the decision enforcing Policy Enforcement Point (PEP)~\cite{b1}. For the sake of simplicity, we will use the term PEP in this paper to represent both mentioned components and their functionalities. The PEP makes and enforces central trust-based Auth* decisions before it decides about the further processing of client's requests. These decisions are based on the trustworthiness of the requestor. Meaning a requestor is only allowed to access the resource, if it's trustworthiness is higher than a predefined trust threshold. The trustworthiness is derived from multiple sources, e.g., user, device and history databases. There can be an arbitrary amount of PEPs in a LAN, all responsible for a specific subset of resources.\par Since of the central positioning of such a PEP component in a ZT architecture, it has access to all information needed to make SFC classification decisions as well. Therefore we propose to combine the functionalities of the SFC classifier and the PEP in the PEP. The PEP can therefore efficiently decide based on their information sources when and which security functions should be applied to packet flows. By doing this, we can achieve three major improvements:
(1) We can dynamically increase the trustworthiness of a requester by applying additional security functions to their packets, or eliminate unnecessary security functions if the trustworthiness after the initial Auth* process is already sufficient to access the requested resource.
(2) We can relieve the PEP of several security functionalities of a ZT network and at the same time distribute these very functionalities by using appropriate service functions that are responsible for this task. 
(3) The Auth* as well as the classification process can be automated by expressing respective deterministic context-based rules in the PEP's policies.

\begin{figure}[h]
    \centering
    \includegraphics[width=.8\columnwidth]{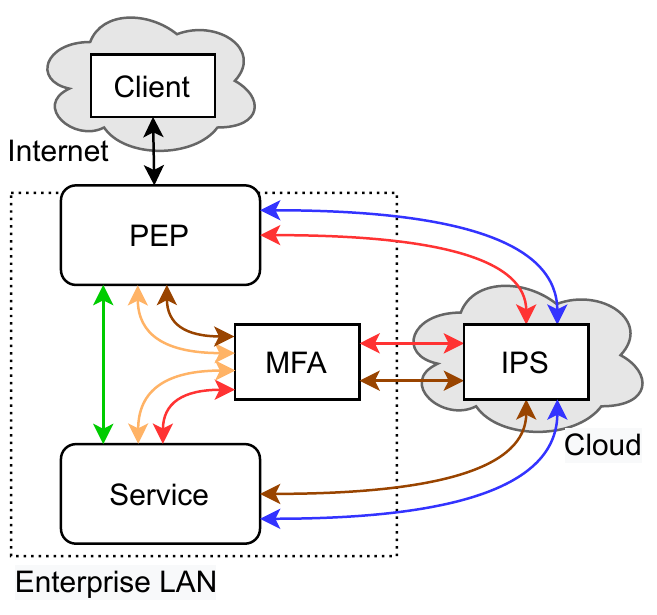}
    \caption{Exemplary ZTSFC and all its chaining options with a PEP as classifier}
    \label{fig:figure1}
\end{figure}

\par Figure~\ref{fig:figure1} depicts a concrete showcase of how the above mentioned improvements can be achieved in practice. Here a client wants to access an enterprise's service. The PEP is the only publicly accessible entry point to this service. All other components are logically micro-segmented by strictly enforcing mutual authentication using enterprise certificates. 
Upon arrival, the client's trustworthiness consists of the presence of a client certificate and management software. If the client can prove that it is a managed device and can present a valid client certificate signed by the PEP, the client's trustworthiness is sufficient and the PEP proxies the client directly to the service (green path). In the case of a device that does only have a valid client certificate but is not managed, the PEP sets up an SFC that includes an IPS function that can partially compensate for the missing security software on the device and increase the trustworthiness of the request to a sufficient level (blue path). In this way, security-related packet processing can be kept efficient by only applying extremely time-consuming tasks such as an IPS if the client is not managed. Only if the IPS is passed successfully, the client's trustworthiness is increased to a sufficient level to be allowed to access the resource. In the case that the device is managed but can not present a valid certificate, the PEP enforces a SFC containing an additional MFA that is performing a check for a different factor (orange path). Leveraging ZTSFCs in such a way unburdens the PEP by outsourcing such security tasks to respective functions. If the client can neither provide a certificate nor is managed, both exemplary functions are part of the enforced SFC (red or brown path; depending on the functions' order). All functions can be set up by the PEP completely dynamically and in an arbitrary order. Furthermore, all these described steps can be automated by defining respective policies. 

\begin{figure}[h]
    \centering
    \includegraphics[width=.8\columnwidth]{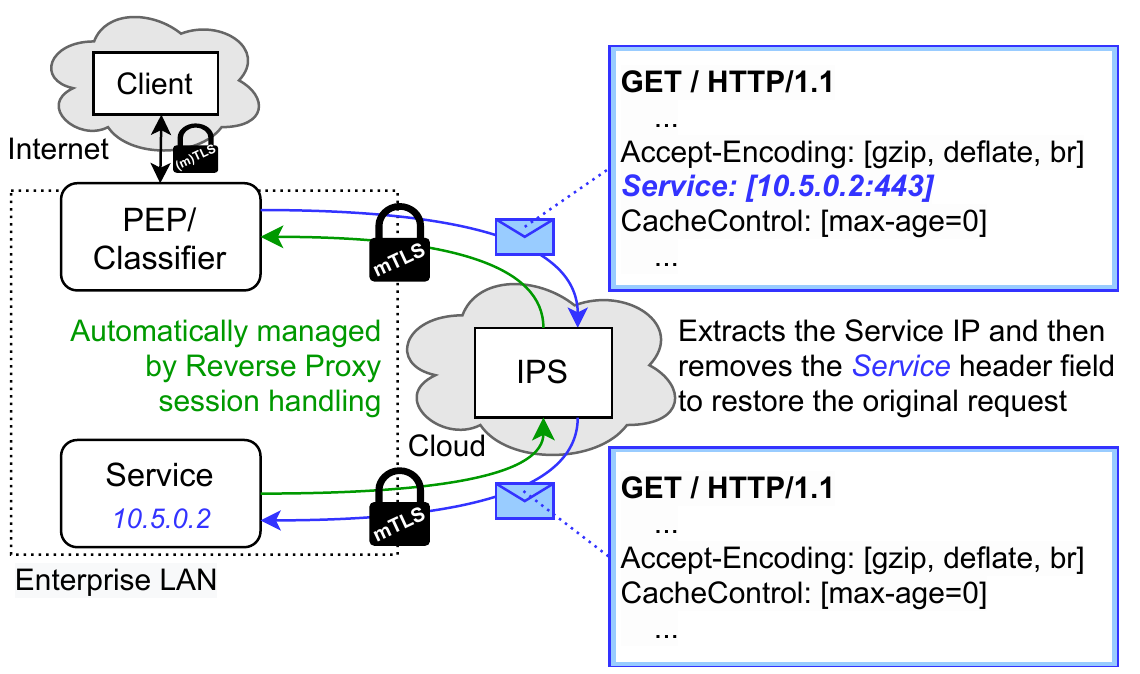}
    \caption{HTTP-based ZTSFC with one target IP (Service) in the HTTP header.}
    \label{fig:figure2}
\end{figure}

\section{HTTP-based ZTSFC}
As a proof of concept, we implemented a first ZTSFC prototype by using TLS and HTTP. We chose this approach because both protocols are widely used and supported by almost every computer system. Thus, this approach has low barriers to entry and can be taken up by anyone easily. \par Figure~\ref{fig:figure2} depicts our early ZTSFC prototype consisting of a PEP, an IPS function and a service. At the reception of a client's request, the PEP extracts all necessary parameters from the packet and decides based on the previously stated context-based deterministic policy rules how to set up the chain. In the case the PEP includes the IPS in the ZTSFC, the PEP adds the respective function IPs into the HTTP header as new custom header field. In this described scenario, it has to add one IP, the IP of the service, to the header. Afterwards, the PEP terminates a mTLS connection with the IPS and proxies the modified HTTP packet to the IPS function. The IPS function can then extract the target's IP of the next hop in the chain from the respective HTTP header fields and proxies it to the service using a new mTLS connection. If the packets do not pass the IPS, they are dropped and the IPS sends respective feedback to the PEP. This approach can be applied analogously to any number of functions. The function IPs are added to the HTTP header similar to an onion routing approach. Each function deletes the IP for it's next hop. Thus, the service receives the original request. The way back through the chain to the PEP is handled by a common session handling similar to a reverse proxy.\par The chosen HTTP approach has shown several advantages. It is completely compliant with the HTTP standard. There is neither a need for a new protocol nor a major adjustment to an existing one. Furthermore, all functions required to implement an HTTP-based ZTSFC can be implemented at the application layer. Thus, the HTTP-based approach is easily realizable, provides a high portability and is independent from lower layer protocols (such as MPLS) that are used to implement SFC technology in other systems. The obligatory confidentiality and integrity are provided by TLS. 

\section{Conclusion \& Future Work}
In this paper, we addressed the problem of inefficient handling of traditional security functions in common ZT architectures. To achieve greater efficiency in this area, we have introduced a novel network security concept that combines the ideas of ZT and SFC. In this context, individual security functions are dynamically orchestrated by the PEP. The concepts leverages context-based deterministic rules evaluated by the Zero Trust PEP to decide which security functions are chained and afterwards applied to the affected packets. By doing this, especially time, power and computing intensive functions such as IPS can be avoided if not necessary. By an automated rule evaluation the whole chaining process can be automated by the PEP. By only using TLS and HTTP this ZTSFC approach can be realized in almost every computer network. Furthermore, the whole approach is TLS and HTTP protocol-compliant. \par
As future work and long-term goal, we plan to set up an elaborated ZTSFC that have well-defined trust-based security policies that enforce highly efficient decisions.
Thereafter, comprehensive performance evaluations of the HTTP approach is scheduled.
Comparisons with other alternative approaches such as MPLS-based ones are also considered.
Furthermore, we are working on additional SFC-related security features. Similar to the onion routing approach, for example, the header field containing the service's IPs could be encrypted with the public key of the respective SFF exchanged during the TLS handshake. By doing this, the other functions in the chain become hidden to the single functions except the adjacent ones. This method would increase the zero trust level in the network. To achieve proof of transit, the functions could add integrity providing security tokens to the HTTP header encrypted by the PEP's public key after the processing of the packet. If a token is absent or shows changes when received by the PEP, it could be dropped.

\section*{Acknowledgment}
This work has been supported in the bwNET2020+ project by the Ministry of Science, Research and the Arts Baden W\"urttemberg (MWK).

\end{document}